\documentclass[twocolumn,prd,nofootinbib,aps,prl,floats,floatfix,amsmath,amssymb,longbibliography,secnumarabic]{revtex4-1} %
\usepackage[english]{babel}
\usepackage{csquotes}
\usepackage{csquotes}

\usepackage[letterpaper,top=2cm,bottom=2cm,left=3cm,right=3cm,marginparwidth=1.75cm]{geometry}

\usepackage{amsmath}
\usepackage{graphicx,slashed}
\usepackage[colorlinks=true,citecolor=red]{hyperref}

\newcommand{\be}{\begin{equation}}
\newcommand{\ee}{\end{equation}}

\newcommand{\bea}{\begin{eqnarray}}
\newcommand{\eea}{\end{eqnarray}}

\begin{document}

\title{Reply to  ``Revisiting bounds on neutrino dark matter interaction at spikes''}
\author{James M.\ Cline}
\email{jcline@physics.mcgill.ca}
\affiliation{McGill University Department of Physics \& Trottier Space Institute, 3600 Rue University, Montr\'eal, QC, H3A 2T8, Canada}
\author{Matteo Puel}
\affiliation{Alperia, via Dodiciville 8, 39100 Bolzano, Italy}

\begin{abstract}
We respond to  \url{https://arxiv.org/pdf/2607.08904} \cite{Farzan:2026wek},
which commented on our previous work to constrain neutrino-dark matter interactions from IceCube observations of neutrinos from active galactic nuclei.

\end{abstract}

\maketitle

Recently Ref.\ \cite{Farzan:2026wek} pointed out some caveats to constraints that we derived on neutrino-dark matter scattering 
\cite{Cline:2022qld,Cline:2023tkp}
using IceCube observations of neutrinos from the active galactic nuclei
(AGNs) TXS 0506+056 and NGC 1068.  The constraints made use of the dense dark matter spikes that are expected to form around the supermassive black hole powering the AGN, which leads to stronger scattering relative to the NFW halo.    We agree with many of the remarks in Ref.\ \cite{Farzan:2026wek}, since we already pointed them out in our original papers.

Let us summarize the caveats that we agree with.  First, it is true that the strong bounds obtained by assuming that the cross section $\sigma(E)$ rises linearly with energy at all energies are not conservative, if in fact $\sigma(E)$ becomes constant at some energy below that which characterizes the AGN emission.  In 
Ref.\ \cite{Cline:2022qld}, we clearly stated our assumptions, and showed how the bounds depend upon the exponent in $\sigma \sim E^n$, for $n=0$ and for $n=1$. 
Moreover in Ref.\ \cite{Cline:2023tkp} we showed how the constraints depend on 
the scale $E_t$ where there is a transition between $n=0$ and $n=1$ in  realistic models, in which $E_t$ depends upon
the DM and mediator masses.  

In Ref.\  \cite{Cline:2023tkp} we  explained that such constraints only apply to models in which all neutrino flavors couple approximately equally, since oscillations are suppressed by matter effects for the interacting flavors.  Hence any noninteracting flavors would escape from the spike unimpeded.  This observation, rediscovered by Ref.\ \cite{Farzan:2026wek}, is not new.

In Ref.\ \cite{Farzan:2026wek}, it was pointed out that the constraints become weak if $\sigma(E)$ decays with energy, rather than becoming constant.  This statement is correct, but we point out that a cross section which decreases with energy is not generic, and can only occur in a renormalizable model if the dark matter 
is a scalar, and if it has a dimensionful coupling $g$ to a scalar mediator.  The reason is that there is always a $t$-channel exchange diagram, which is infrared sensitive to the mediator mass $m$ in the phase space integration.   
This leads to $1/m^2$ dependence in the cross section, which already has the correct dimensions to be a cross section.
Only if there is a dimensionful factor $g^2$ in the numerator can $\sigma(E)\sim 1/s$, in terms of the Mandelstam variable $s$.  For dimensionless couplings to the mediator, $\sigma(E)\sim m_\chi E/m^2$ for $E < m^2/m_\chi$, where $m_\chi$ is the DM mass.  The tree-level cross section tends toward  constant behavior  at higher energies.  In general, 
the growth with energy is only limited by the Froissart bound $\sigma(E)\sim \ln^2(s)$.

Next we comment on some points in Ref.\ \cite{Farzan:2026wek} which we believe to be incorrect.  First, the expression given there for the cross section with the $Z'$ mediator (above Eq.\ (2)) seems to assume scattering of identical fermions, which has an interfering $u$-channel contribution causing the cross section to fall as $1/s$.  But the dark matter is not neutrinos, so there is never a $u$-channel diagram, and our previous argument leading to a constant 
cross section applies.\footnote{Moreover the expression (2) in Ref.\ \cite{Farzan:2026wek} for $d\sigma/dE$ does not have the correct dimensions.}  As noted above, a cross section falling with $s$ leads to weak limits, but this is not the appropriate case to consider.

A further criticism made by Ref.\ \cite{Farzan:2026wek} is that it is difficult to circumvent the annihilation plateau that cuts off the dark matter spike at small radii.  We pointed out that asymmetric dark matter automatically avoids this caveat.  Ref.\ \cite{Farzan:2026wek} notes that the spike then creates an effective mass for the neutrinos which suppresses their oscillations.  In fact, even for symmetric dark matter, there is an energy-dependent contribution to the neutrino self-energy
that becomes comparable to the usual matter effect when $E_\nu\gtrsim E_t \sim m^2/m_\chi$ \cite{Notzold:1987ik}, which is in the regime of constant cross section.  
The issue of neutrino oscillations being damped in the DM spike has already been discussed above.

\bibliographystyle{utphys}
\bibliography{sample}

@article{Farzan:2026wek,
    author = "Farzan, Yasaman",
    title = "{Revisiting bounds on neutrino dark matter interaction at spikes}",
    eprint = "2607.08904",
    archivePrefix = "arXiv",
    primaryClass = "hep-ph",
    month = "7",
    year = "2026"
}

@article{Notzold:1987ik,
    author = {N{\"o}tzold, Dirk and Raffelt, Georg},
    title = "{Neutrino dispersion at finite temperature and density}",
    reportNumber = "MPI-PAE/PTh-87/87",
    doi = "10.1016/0550-3213(88)90113-7",
    journal = "Nucl. Phys. B",
    volume = "307",
    pages = "924--936",
    year = "1988"
}

@article{Cline:2022qld,
    author = "Cline, James M. and Gao, Shan and Guo, Fangyi and Lin, Zhongan and Liu, Shiyan and Puel, Matteo and Todd, Phillip and Xiao, Tianzhuo",
    title = "{Blazar Constraints on Neutrino-Dark Matter Scattering}",
    eprint = "2209.02713",
    archivePrefix = "arXiv",
    primaryClass = "hep-ph",
    doi = "10.1103/PhysRevLett.130.091402",
    journal = "Phys. Rev. Lett.",
    volume = "130",
    number = "9",
    pages = "091402",
    year = "2023"
}

@article{Cline:2023tkp,
    author = "Cline, James M. and Puel, Matteo",
    title = "{NGC 1068 constraints on neutrino-dark matter scattering}",
    eprint = "2301.08756",
    archivePrefix = "arXiv",
    primaryClass = "hep-ph",
    doi = "10.1088/1475-7516/2023/06/004",
    journal = "JCAP",
    volume = "06",
    pages = "004",
    year = "2023"
}

\end{document}